\documentstyle[12pt]{article}
\def\al{\alpha}

\def\ga{\gamma}
\def\de{\delta}
\def\ep{\epsilon}

\def\th{\theta}

\def\ph{\phi}

\def\ps{\psi}
\def\om{\omega}
\def\Ga{\Gamma}
\def\De{\Delta}

\def\Ps{\Psi}

\def\fr#1#2{{{#1} \over {#2}}}

\def\vev#1{\langle {#1}\rangle}
\def\bra#1{\langle{#1}|}
\def\ket#1{|{#1}\rangle}

\def\half{{\textstyle{1\over 2}}}
\def\frac#1#2{{\textstyle{{#1}\over {#2}}}}

\def\lsim{\mathrel{\rlap{\lower4pt\hbox{\hskip1pt$\sim$}}
    \raise1pt\hbox{$<$}}}
\def\gsim{\mathrel{\rlap{\lower4pt\hbox{\hskip1pt$\sim$}}
    \raise1pt\hbox{$>$}}}
\def\sqr#1#2{{\vcenter{\vbox{\hrule height.#2pt
         \hbox{\vrule width.#2pt height#1pt \kern#1pt
         \vrule width.#2pt}
         \hrule height.#2pt}}}}

\newcommand{\beq}{\begin{equation}}
\newcommand{\eeq}{\end{equation}}
\newcommand{\bea}{\begin{eqnarray}}
\newcommand{\eea}{\end{eqnarray}}
\newcommand{\rf}[1]{(\ref{#1})}

\renewenvironment{thebibliography}[1]
 { \rm
   \begin{list}{\arabic{enumi}.}
    {\usecounter{enumi} \setlength{\parsep}{0pt}
     \setlength{\itemsep}{3pt} \settowidth{\labelwidth}{#1.}
     \sloppy
    }}{\end{list}}

\begin{document}
\titlepage

\begin{flushright}
{COLBY-93-05\\}
{IUHET 256\\}
\end{flushright}
\vglue 1cm

\begin{center}
{{\bf ATOMIC SUPERSYMMETRY, RYDBERG WAVE PACKETS,
\\}
\vglue 0.2cm
{\bf AND RADIAL SQUEEZED STATES
\\}
\vglue 1.0cm
{Robert Bluhm$^a$ and V. Alan Kosteleck\'y$^b$\\}
\bigskip
{\it $^a$Physics Department\\}
\medskip
{\it Colby College\\}
\medskip
{\it Waterville, ME 04901, U.S.A\\}
\bigskip
{\it $^b$Physics Department\\}
\medskip
{\it Indiana University\\}
\medskip
{\it Bloomington, IN 47405, U.S.A.\\}

\vglue 0.8cm
}
\vglue 0.3cm

\end{center}

{\rightskip=3pc\leftskip=3pc\noindent
We study radial wave packets produced by short-pulsed laser fields
acting on Rydberg atoms,
using analytical tools
from supersymmetry-based quantum-defect theory.
We begin with a time-dependent perturbative
calculation for alkali-metal atoms,
incorporating the atomic-excitation process.
This provides insight into the general wave packet behavior
and demonstrates agreement with conventional theory.
We then obtain an alternative analytical description of
a radial wave packet as a member of a particular family
of squeezed states,
which we call radial squeezed states.
By construction,
these have close to minimum uncertainty in the radial coordinates
during the first pass through the outer apsidal point.
The properties of radial squeezed states are investigated,
and they are shown to provide a description of
certain aspects of Rydberg atoms
excited by short-pulsed laser fields.
We derive expressions for the time evolution
and the autocorrelation of the radial squeezed states,
and we study numerically and analytically their behavior
in several alkali-metal atoms.
Full and fractional revivals are observed.
Comparisons show agreement
with other theoretical results and with experiment.

}

\vskip 0.3truein
\centerline{\it Published in Physical Review A
{\bf 49}, 4628 (1994)}

\vfill
\newpage

\baselineskip=20pt
{\bf\noindent I. INTRODUCTION}
\vglue 0.4cm

When an atom is excited from its ground state
by a short-pulsed laser field,
a localized radial wave packet is produced
that has behavior mimicking the
classical radial motion of a charged particle in a Coulomb field
\cite{pasa,pasb,alrz,alz}.
The properties and time evolution of such wave packets
provide interesting experimental and theoretical opportunities
to probe the interface between classical and quantum mechanics.

Experiments have detected the periodic motion of the electron in
a Rydberg atom excited by a short laser pulse,
with period equal to the classical period of a particle
in a keplerian orbit
\cite{wnlh,ymps}.
The motion of the wave packet is only partly classical,
however.
For excitations by a single laser pulse
from the atomic ground state,
the angular distribution is that of a p state.
Also, the radial wave packet disperses over the course of time.
However,
after many Kepler times,
the wave packet remnants eventually recombine
into a packet that is close to the original shape
and that oscillates radially with the keplerian orbital period.
This recombined wave packet is called a full revival.
In the intervals between the initial and the fully revived
classical motions,
subsidiary wave packets form.
These are called fractional revivals,
and they have orbital periods equal to rational fractions of the
classical keplerian period.
Both full and fractional revivals
have been seen experimentally
\cite{yms,yes,mmhe},
and descriptions of their behavior have been developed
\cite{wnmh,avp,naa}.

Several theoretical approaches have been used to study the
properties of radial wave packets formed by
excitation of Rydberg atoms with short laser pulses.
Refs.\ \cite{pasa,pasb}
solve the Schr\"odinger equation numerically,
while
for gaussian shaped laser pulses and weak-field excitations,
Refs.\ \cite{alrz,alz}
use perturbation theory.
For a particular class of pulse shapes,
Ref.\ \cite{grl}
is able to obtain nonperturbative solutions.
Also,
Ref.\ \cite{ndg}
provides a theoretical description for the generation
of Ramsey fringes with Rydberg wave packets.

Since the radial wave packets
are initially localized in the radial
coordinates and partly follow the classical motion,
a description in terms of some kind of coherent state
\cite{kls}
might seem appropriate.
However,
standard coherent-state approaches
to the hydrogenic Coulomb problem
\cite{schr,b,mos,ni,mcab}
and to short-pulse laser excitation of Rydberg atoms
\cite{nab,gadb,dags}
have yielded descriptions
of the motion in a Coulomb potential
with the electron moving on circular or elliptical orbits.
None of these match the behavior of Rydberg atoms in
short-pulsed laser fields
with p-state angular distributions.
Within this framework,
the issue of an analytical construction of radial wave packets
for Rydberg atoms prepared by excitation with
short laser pulses
with no external fields present has been an open problem.

Recently,
we have provided a framework for an analytical study
of Rydberg wave packets
\cite{bkrss}
and have discussed some of their properties
for hydrogen.
In the present work,
we generalize this method to non-hydrogenic atoms,
with particular emphasis on the alkali-metal atoms
for which experiments have been performed.
We also extend our analysis and
obtain further results for the hydrogenic case.

To incorporate non-hydrogenic features in the treatment,
we address matters in the context of
supersymmetry-based quantum-defect theory (SQDT)
\cite{kona,konb}.
This theory describes
the behavior of an excited Rydberg electron as that
of a single particle in an effective central potential.
The effective potential can be found by acting on
the hydrogenic Coulomb potential with a symmetry transformation
and adding a symmetry-breaking term.
The symmetry transformation used
is a quantum-mechanical supersymmetry
\cite{niw}.
The supersymmetry-breaking term incorporates
nonhydrogenic contributions to the effective potential.
Although the model is analytical,
its combination of supersymmetry
ideas with the notion of quantum defects
\cite{ryd,qdta,qdtb,sim}
is such that exact energy eigenvalues are generated.
The corresponding analytical eigenfunctions
can be used to provide predictions for
transition probablilities of alkali-metal atoms
\cite{konb}
and alkaline-earth-metal ions
\cite{dje},
in good agreement with accepted values.
The model also can be used to generate
Stark maps for the alkali-metal atoms
\cite{blk}
that agree with experiment and standard theory
\cite{klep,gall}.
The validity of the model breaks down inside the electronic core,
although some bulk features of the fine structure of
alkali-metal atoms are reproduced
\cite{knta}.
Extensions of these ideas apply to other situations
including, for example, the Penning trap
\cite{aka}.
For recent overviews of atomic supersymmetry,
see Ref.\ \cite{akb}.

The analytical inclusion of quantum defects
for Rydberg atoms suggests that SQDT could be a
useful tool for the study of
Rydberg atoms in short-pulsed laser fields.
We have identified two approaches as being
especially promising in this regard.
One is the inclusion of SQDT in conventional
time-dependent perturbation theory
to explore the formation and behavior of radial wave packets,
in particular to show the presence of oscillatory motion
between the orbital apsidal points.
The results can then be compared with those obtained using
numerical approaches
\cite{pasa,pasb}
and WKB solutions in perturbation theory
\cite{alrz,alz}.

The second approach we adopt is the use of SQDT
in the construction of a family of analytical squeezed states,
which we call radial squeezed states (RSS),
that can represent radial wave packets
formed in a Rydberg atom excited by a short laser pulse.
For our purposes,
given a hamiltonian and a commutation relation,
a squeezed state can be taken as a wave function
that satisfies the corresponding minimum-uncertainty relation
at a particular time but with uncertainties
not necessarily those of the ground state.
For the simple harmonic oscillator,
for example,
the uncertainty product of a squeezed state has a
sinusoidal dependence on time,
with the uncertainties in coordinate and momentum space
oscillating while the center of the wave packet
follows the classical motion
\cite{niss}.

The analytical method we use to obtain the RSS
is discussed in Sec.\ IV.
The result is a three-parameter family of wave packets
with the angular dependence of a p state.
For an individual packet,
the values of the three parameters are fixed
in terms of expectations of the energy, momentum,
and position when the wave packet first reaches
the outer apsidal point.
This initialization condition agrees both with the
results of other theoretical calculations
and with the
intuitive notion that the electron wave packet
attains closest-to-classical behavior when localized
at large quantum numbers immediately after its formation.
The ensuing behavior is completely determined
by the time-dependent Schr\"odinger equation.

Since the RSS are given by a relatively simple
analytical expression,
their time evolution can be studied
either analytically or numerically.
We have used both approaches.
Quantities of interest are the time dependence
of the RSS and of autocorrelation functions with the initial packet.
In what follows,
we examine these for several different alkali-metal atoms.
We treat issues such as the variation with the quantum defect
of the orbital period and of the full or fractional revivals.

This paper is organized as follows.
In the next section,
we summarize some necessary background information
on the salient features of SQDT
and on the various approaches
to the study of radial wave packets
formed with short-pulsed laser fields applied to Rydberg atoms.
In Sec.\ III, we use SQDT to perform a time-dependent
perturbative calculation for the formation and time evolution
of the radial wave packets.
The derivation and basic properties of the RSS are described
in Sec.\ IV.
Section V discusses their time-dependent properties,
including their time evolution
and their autocorrelation functions,
comparing with other available
theoretical results and with some experiments.
A summary and our conclusions are given in Sec.\ VI.
Throughout this work,
we use atomic units with $\hbar=e=m_e=1$.

\vglue 0.6cm
{\bf\noindent II. BACKGROUND}
\vglue 0.4cm

In this section,
we provide background material on atomic supersymmetry
and on prior theoretical treatments of Rydberg
atoms in short-pulsed laser fields.

\vglue 0.6cm
{\bf\noindent A. Atomic Supersymmetry}
\vglue 0.4cm

Atomic supersymmetry involves
a particular realization $\cal{R}$
of the superalgebra sqm(2).
The three generators of sqm(2) can be taken as the
hamiltonian $H_{ss}$ of the supersymmetric quantum system,
the supersymmetry charge $Q$,
and its conjugate $Q^\dagger$.
They obey the relations
\beq
[H_{ss},Q]=[H_{ss},Q^{\dag}]=0~~,~~~~
\{Q,Q^{\dag} \} = H_{ss}
\quad .
\label{acr}
\eeq
In the realization $\cal{R}$,
the hamiltonian for sqm(2) splits as the direct sum
$H_{ss} = H_+ \oplus H_-$.
The component hamiltonians $H_+$ and $H_-$ are one-variable
differential operators satisfying the eigenequations
\beq
H_\pm \Psi_{\pm n} \equiv \bigl[ -{{d^2} \over {dx^2}} +
V_\pm (x) \bigr] \Psi_{\pm n} = \epsilon_n \Psi_{\pm n}
\quad ,
\label{hpm}
\eeq
where the potentials $V_+$ and $V_-$ are supersymmetric
partners conventionally expressed in terms of
$x$-derivatives of a function $U(x)$:
\beq
V_\pm (x) = (\half U^\prime )^2 \mp \half U^{\prime\prime}
\quad .
\label{hpma}
\eeq
The ground-state eigenvalue is zero and
is associated with $H_+$ only.
The spectra of the two hamiltonians are otherwise identical
\cite{niw}.

When the hamiltonian $H_+$ is identified with
the differential operator for the
radial part of the hydrogen-atom Schr\"odinger equation
expressed in spherical polar coordinates,
the supersymmetry partner $H_-$
can be constructed
\cite{kona}
by fixing $l$ and solving for $V_{-l}$ in
Eq.\ \rf{hpm}.
It can then be shown that
the supersymmetric eigenfunction partners are
$R_{nl}$ with $n \geq 1$
and $R_{n,l+1}$ with $n \geq 2$.
A physical interpretation of these results
has been proposed
\cite{kona},
leading to sets of nested supersymmetries interconnecting
eigenvalues and eigenstates of different atoms and ions.
For example,
in the exact symmetry limit
where electron-electron interactions can be neglected,
the s levels of lithium may be regarded
as the supersymmetric partners of the hydrogen atom s levels.
The s orbitals of sodium can then in turn be viewed as
supersymmetric partners of the lithium s orbitals.
Connections also exist among p and higher orbitals.
An integer shift in the angular quantum number $l$
accompanies each supersymmetry operation.

In real atoms such as alkali metals,
the exact supersymmetry is broken by
coupling between the valence electron
and the electronic core.
One result is a shift in eigenenergies
\cite{ryd},
$E_n \rightarrow E_{n^{\ast}} = -1/2n^{\ast 2}$,
relative to hydrogenic values.
The quantity
$n^{\ast} = n - \delta (n,l)$
incorporates these shifts via
quantum defects $\delta (n,l)$.
As $n$ increases,
the exact quantum defects
rapidly approach asymptotic values
$\delta (l)$.

The basic idea of SQDT is to incorporate these shifts
in an analytical one-particle hamiltonian
\cite{konb}.
The model is obtained by adding
to the exact supersymmetric hamiltonian
specific supersymmetry-breaking terms, chosen
so that the quantum-defect eigenenergies are reproduced
while leaving the eigenfunctions analytical.
A modified angular quantum number
$l^{\ast} = l - \delta(l) + I(l)$
is introduced,
where $I(l)$ is an integer playing the role of the
supersymmetric shift.
The SQDT hamiltonian is found via the replacement
$n,l,E_n \rightarrow n^{\ast},l^{\ast},E_{n^{\ast}}$
in the radial Coulomb equation.
The corresponding SQDT eigenfunctions
are given by
\beq
R_{{n^{\ast}} {l^{\ast}}}(r) = {2\over{{n^{\ast}}^2}}
 \Bigl[ {{\Gamma({n^{\ast}}-{l^{\ast}})}
 \over{\Gamma({n^{\ast}}+{l^{\ast}}+1)}} \Bigr]^{1\over2}
 \Bigl({{2r}\over {n^{\ast}}} \Bigr)^{l^{\ast}} \exp
 \Bigl( -{r\over {n^{\ast}}} \Bigr)
 L_{{n^{\ast}}-{l^{\ast}}-1}^{(2{l^{\ast}}+1)} \Bigl( {{2r}
 \over {n^{\ast}}} \Bigr)
\quad .
\label{rdef}
\eeq
The full three-dimensional SQDT wave functions
are therefore
$R_{{n^{\ast}} {l^{\ast}}}(r) Y_{l m}(\th,\ph)$.
For asymptotic quantum defects  $\delta (l)$,
these eigenfunctions form a complete and orthonormal set.
Note that in the limit of vanishing quantum defects $\de (l)$
and supersymmetry integers $I(l)$
the usual Coulomb eigenvalues and eigenfunctions are recovered.

\vglue 0.6cm
{\bf\noindent B. Radial Wave Packets}
\vglue 0.4cm

The time-dependent Schr\"odinger equation
for a hydrogen atom in a laser field is
\beq
i \fr \partial {\partial t} \Psi (t) =
(H_0 - {\vec \mu} \cdot {\vec E}) \Psi (t)
\quad ,
\label{se}
\eeq
where $H_0$ is the zero-field hamiltonian,
${\vec \mu}$ is the atomic dipole moment,
and ${\vec E}$ is the laser field.
Consider an atom excited by a single photon from an
initial state $\ket i $ to a range of Rydberg states by a
laser pulse of length $\tau$ and frequency $\om$.
The electric field can be written as
\beq
{\vec E} (t) = {\cal E} (t) \, {\vec \ep} \, e^{-i \om t}
+ {\rm c.c.}
\quad .
\label{efield}
\eeq
Here,
${\vec \ep}$ is the polarization vector and
${\cal E} (t)$ is a gaussian envelope for the laser pulse,
\beq
{\cal E} (t) = {\cal E}_0~
\exp\left( {- \fr {4 (\ln 2 ) t^2} {\tau^2}}\right)
\quad ,
\label{env}
\eeq
which is taken to be centered at $t=0$
with full width at half maximum equal to $\tau$.

If we expand $\Psi (t)$ in terms of hydrogenic wave functions,
\beq
\Psi (t) = \sum_{nlm} a_{nlm} (t) e^{-i E_n t}R_{nl}Y_{lm}
\quad ,
\label{expans}
\eeq
and use the rotating wave approximation,
we obtain equations for the time-dependent
parameters
$a_{nlm} (t)$.
These equations may be solved numerically
\cite{pasa,pasb}
or perturbatively for weak fields
\cite{alrz,alz}.
Quantities of interest,
such as the probability distribution
$r^2 | \Psi (r,t) |^2$
of the radial part of the wave function,
may then be plotted as a function of time.
Refs.\ \cite{pasa,pasb,alrz,alz}
use hydrogenic dipole moments and energy spacings
for simplicity,
although numerical methods incorporating standard
quantum-defect theory can be applied.
Refs.\ \cite{ymps,yms,yes}
contain numerical calculations for
alkali-metal atoms.

The results obtained by these procedures show that a radial
wave packet forms shortly after the initial laser excitation.
The packet subsequently
oscillates between the radial apsidal points.
Refs.\ \cite{pasa,pasb,alrz,alz}
consider an 8 to 10 psec laser pulse that
excites a range of p states of hydrogen
centered about the value $\bar n=85$ of $n$.
The resulting wave packet
is calculated to oscillate with a period
equal to the classical Kepler period
$T_{\rm cl} = 2 \pi \bar n^3 \simeq 93.3$ psec.
At the inner turning point,
near the origin,
the distribution $r^2 | \Psi (r,t) |^2$
of the packet is broad and exhibits many oscillations.
As the packet returns to the
outer turning point,
near $2 \bar n^2=14450$ a.u.,
the distribution becomes narrower and increases in amplitude,
while the number of oscillations decreases significantly.
At the first pass through the outer turning point,
the wave packet is calculated to have
close to minimum uncertainty in the radial coordinates
\cite{pasa,pasb}.

\vglue 0.6cm
{\bf\noindent III. TIME-DEPENDENT PERTURBATION THEORY}
\vglue 0.4cm

In this section,
we show how to perform a time-dependent perturbative
calculation that incorporates
SQDT into the analysis.
The calculation describes
the formation and time evolution of a Rydberg
atom excited from its ground state by a short laser pulse.
The field is taken to have the form in
Eq.\ \rf{efield}
with envelope ${\cal E} (t)$ as in
Eq.\ \rf{env}.

We begin by expanding
the wave function $\Psi (t)$
describing alkali-metal Rydberg states
in terms of the complete
and orthonormal set of SQDT states
$R_{{n^{\ast}} {l^{\ast}}}(r) Y_{l m}(\th,\ph)$.
For single photon excitations from the ground state,
only p-state angular wave functions are needed.
However,
the procedure that follows can be generalized
to multi-photon processes that excite d or higher states.
We neglect any contributions from the continuum,
which are negligible for the cases of interest.
The expansion \rf{expans} therefore becomes
\beq
\Psi (t) = Y_{1 0} (\th,\ph) \sum_n a_n (t)
\, e^{-i E_{n^{\ast}} t} \, R_{{n^{\ast}} {l^{\ast}}}(r)
\quad ,
\label{susyexps}
\eeq
where now $n^\ast = n - \de (1)$ and
$l^\ast = 1 - \de (1) + I(1)$,
with $\de (1)$ the p-state quantum defect and
$I(1)$ the p-state supersymmetry integer.
We then use the rotating wave approximation
and solve for $a_n (t)$ to lowest order in
perturbation theory.
For times $t \gg 0$,
we get
\beq
a_n =  i \bra {n^\ast, l^\ast}
{\vec \mu} \cdot {\vec \ep} \ket i
{\tilde {\cal E}} (\de_n)
\quad ,
\label{acoeff}
\eeq
where $ \ket i$ denotes the ground state
and ${\tilde {\cal E}} (\de_n)$ is the Fourier transform
of ${\cal E} (t)$
with $\de_n = (E_{n^\ast} - E_{i^\ast} - \om)$.
Note that the eigenenergies $E_{n^\ast}=-1/2n^{\ast 2}$,
including the ground-state energy $E_{i^\ast}$,
incorporate the quantum defect.

The dipole matrix elements
$\bra {n^\ast, l^\ast} {\vec \mu} \cdot {\vec \ep} \ket i$
can be evaluated analytically
for different alkali-metal atoms
\cite{konb}.
Equation \rf{rdef}
specifies the form of the radial wave functions
$R_{{n^{\ast}} {l^{\ast}}}(r)$.
Then,
for fixed $\th$ and
up to a normalization constant $N$,
the radial probability density is given by
\beq
r^2 {| \Psi (t) |}^2 = N r^2 \Bigm\vert
\sum_n R_{{n^{\ast}} {l^{\ast}}}(r)~
\bra {n^\ast, l^\ast} r \ket {1, 0}~
\exp\left( { - \fr {(E_{n^\ast} - {E_{{\bar n}^\ast}})^2 \tau^2}
{16 \ln 2}}\right)~
e^{- i E_{n^\ast} t} \Bigm\vert^2
\quad .
\label{rprob}
\eeq
In this expression,
we have taken the laser frequency $\om$
as tuned to excite a range of energies
centered on $E_{{\bar n}^\ast}$,
\beq
\om = E_{{\bar n}^\ast} - E_{i^\ast}
\quad .
\label{omg}
\eeq

Equation \rf{rprob} holds for $t \gg 0$.
We can modify the procedure and evaluate
$r^2 {| \Psi (t) |}^2$ for $t=0$ as well.
The resulting calculation shows that the
answer can be written by making the substitution
\beq
e^{- i E_{n^\ast} t}  \rightarrow
\half \Bigl[ 1 + {\rm erf} \Bigl( -i
\sqrt{\fr {(E_{n^\ast} - {E_{{\bar n}^\ast}})^2 \tau^2} {16 \ln 2}}
\Bigr) \Bigr]
\quad
\label{sub}
\eeq
in Eq.\ \rf{rprob},
where erf(z) is the error function.

As a first example,
consider hydrogen.
This element has $\de (1) = I(1) = 0$,
so $n^\ast = n$.
For purposes of comparison with
Ref.\ \cite{alrz},
we take $\bar n^\ast ={\bar n} = 85$ and $\tau = 8$ psec.
Figure 1 displays the radial probability distribution
$f(r) = r^2 {| \Psi (t) |}^2$ at several different times.
{}From left to right,
these are
$t =
 \fr 1 9 T_{\rm cl}\simeq 10.3$ psec,
$\fr 2 9 T_{\rm cl}\simeq 20.7$ psec,
$\fr 3 9 T_{\rm cl}\simeq 31.1$ psec, and
$\fr 4 9 T_{\rm cl}\simeq 41.5$ psec,
where $T_{\rm cl} = 2 \pi {\bar n}^3\simeq 93.3$ psec
is the classical orbital period.
No vertical scale is shown
because the packet is unnormalized.
The figure uses
Eq.\ \rf{rprob}
evaluated numerically over the subset of states
$n = 75$ -- $95$,
which we determined to be sufficient for high accuracy.

The graph reveals that
during formation the radial wave packet
moves away from the origin,
increasing in amplitude as it approaches
the outer turning point
$r_{\rm cl} \approx 2n^2 = 14450$ a.u.
It then decreases in amplitude
as it starts to bounce back towards the origin.
Our Fig.\ 1 agrees with
Figure 1 of Ref.\ \cite{alrz}
in its gross features,
although there are differences
in the detailed shapes of the packets.
Ref.\ \cite{alrz}
uses a WKB solution to evaluate the Coulomb functions,
whereas we use
Eq.\ \rf{rprob}
directly and evaluate the associated Laguerre polynomials
in $R_{{n^{\ast}} {l^{\ast}}}(r)$
with high numerical accuracy for a subset of states.

As a second example,
we look at rubidium.
This has $\de (1)\simeq 2.65$ and $I(1) = 3$.
For purposes of comparison with the hydrogenic case,
we again choose $\tau = 8$ psec and ${\bar n} = 85$.
This corresponds to ${{\bar n}^\ast} \simeq 82.35$.
As we show in Sec.\ IV,
the inequivalence of $\bar n$ and $\bar n^\ast$
means that the classical orbital period
is reduced from the hydrogenic value to
$T^*_{\rm cl} = 2 \pi {\bar n}^{\ast 3}\simeq 84.9$ psec.
Figure 2 displays the radial probability distribution $f(r)$
for rubidium at the same times as in Fig.\ 1.
The overall behavior is similar to that of hydrogen.
However,
comparison of
Figs. 1 and 2 shows that
the outer turning point $r^\ast_{\rm cl}$ for rubidium
is slightly smaller than for hydrogen.
This difference is due to the quantum defects.
In fact,
as we show in Sec.\ IV A,
the outer turning point is
$r^\ast_{\rm cl} \approx 2 \bar n^{\ast 2}
\simeq 13560$ a.u. here.

\vglue 0.6cm
{\bf\noindent IV. RADIAL ATOMIC SQUEEZED STATES}
\vglue 0.4cm

The results in the previous section demonstrate
that the analytical framework provided by SQDT
is suitable for the perturbative study
of the time evolution of radial wave packets
for different alkali-metal atoms.
However,
the development and application of
an analytical treatment of the problem that
avoids perturbation theory is evidently of interest
and is the principal goal of this paper.

The uncertainty product
$\De r \De p_r$ is close to its minimum value
during the first passage of the wave packet
through the outer turning point
\cite{pasa,pasb}.
This suggests that we seek an analytical description of
a radial wave packet at its outer turning point
as some type of minimum-uncertainty wave function.

In this section,
we use SQDT to derive a family of analytical wave packets
that obey minimum-uncertainty relations for
certain special variables
and that have uncertainty product
$\De r \De p_r$ close to the minimum value.
We call these packets
radial squeezed states (RSS).
The use of SQDT ensures that the RSS
are relevant for all alkali-metal atoms,
not merely hydrogen,
and permits us to
take into account the quantum defects of the different
alkali-metal atoms used in experiments.
Discussion of the time evolution of the RSS
is deferred to Sec.\ V.

\vglue 0.6cm
{\bf\noindent A. Classical Motion}
\vglue 0.4cm

The construction of close-to-classical
(approximate minimum-uncertainty) wave functions is aided
by an understanding of classical behavior.
In this subsection,
we consider the motion of a classical particle in a
central potential that can be viewed as the
classical limit of the effective potential in SQDT.

In general,
the motion of a particle in a central potential $U(r)$ is
governed by the lagrangian
\beq
L = \half {\dot r}^2 + \half r^2 {\dot \th}^2 - U(r)
\quad .
\label{ham}
\eeq
The classical angular momentum
$l= r^2 {\dot \th}$
is constant
and so can be used to reduce the equations of motion
to a single equation for the radial coordinate $r$.

The SQDT uses an $l$-dependent effective
central potential $U(r)$.
One perspective is therefore that it
replaces the intractable many-body action
for the multi-electron atom
with an infinite set
of single-particle effective actions,
labeled by $l$.
Each member of this set
describes the behavior of the system for
the given value of the angular momentum.
In the classical case,
$l$ is a continuous variable,
and the central potential $U(r)$
can be taken as
\beq
U(r) = - \fr 1 r + \fr {\l^{\ast 2} - l^2} {2r^2}
\quad .
\label{udef}
\eeq
This potential generates an effective radial hamiltonian
\beq
H^\ast \equiv
\half p_r^2 + \fr {l^{\ast 2}} {2 r^2} - \fr 1 r
= E^\ast
\quad ,
\label{energy}
\eeq
where $p_r = \dot r$ is the radial momentum
and $E^\ast$ is the energy.
Note that when $l^\ast = l$,
the classical SQDT reduces to the usual Coulomb case.

For negative $E^\ast$,
the radial motion is oscillatory between two values
$r_{1,2}$ of $r$
corresponding to the apsidal points of the orbit
and given by
\beq
r_{1,2} = \fr 1 {2 \vert E^\ast\vert}
\left( 1 \pm \sqrt{1 - 2 \vert E^\ast \vert l^{\ast 2}}
\right)
\quad .
\label{rclE}
\eeq
The classical orbital period
$T^\ast_{\rm cl}$ for the SQDT
can be defined as the time taken to move
from $r_1$ to $r_2$ and back.
Using Eq.\ \rf{energy},
a short calculation gives
\beq
T^\ast_{\rm cl} = \fr {\pi} {\sqrt{2\vert E^\ast \vert ^3}}
\quad .
\label{tclqd2E}
\eeq

Bertrand's theorem implies that for the
SQDT central potential \rf{udef}
the orbits are not closed.
However,
the classical equations of motion can be
solved exactly,
as might be expected from the analytical nature of the
quantum theory.
The solution for the classical orbit
can be found by eliminating $t$
from
Eq.\ \rf{energy},
using $d\th = l dt/r^2$,
and integrating.
We thereby obtain the orbit equation
\beq
\fr 1 r = \fr 1 {l^{\ast 2}}
\left( 1 + e \cos [f(\th - \th_0)] \right)
\quad .
\label{orbit}
\eeq
Here,
$\th_0$ is a constant of the integration,
and we have defined
\beq
f = \fr {l^\ast} l
\quad
\label{fratio}
\eeq
and
\beq
e = \sqrt{1 - 2 \vert E^\ast\vert l^{\ast 2} }
\quad .
\label{ecc}
\eeq

Equation \rf{orbit} describes
a precessing ellipse,
with semimajor axis
$a=1/2\vert E^\ast \vert$
and eccentricity $e$.
The apsidal points of the orbit are
$r_{1,2} = a (1 \pm e)$.
The precession in one orbital period is
$\De \th = 2 \pi (1- 1/f)$.
Evidently,
the precession is a classical consequence
of the presence in the SQDT
of the supersymmetry-type shifts
and the quantum defects.
Note that for different atomic levels
the quantity $f$ in the SQDT
can be larger or smaller than one,
so the precession may be clockwise or
counterclockwise.
Also,
if $l^\ast = l$
then $f=1$ and the precession disappears.

For convenience in comparing with the quantum theory,
it is useful to write $E^\ast$ as
\beq
E^\ast = -\fr 1 {2 n^{\ast 2}}
\quad .
\label{nstar}
\eeq
The choice of notation reflects that
in the quantum theory
$n^\ast$ is
the quantized, shifted principal quantum number.
However, at the classical level,
it is a convenient, continuous variable.
The classical turning points
and the classical orbital period
can be expressed in terms of $n^\ast$ as
\beq
r_{1,2} = n^{\ast 2} \left( 1 \pm \sqrt{1 - \fr {l^{\ast 2}}
{n^{\ast 2}}} \right)
\quad
\label{rcl}
\eeq
and
\beq
T^\ast_{\rm cl} = 2 \pi n^{\ast 3}
\quad .
\label{tclqd2}
\eeq
The orbital eccentricity becomes
\beq
e = \sqrt{ 1 - \fr {l^{\ast 2}} {n^{\ast 2}} }
\quad .
\label{eccnstar}
\eeq
For single-photon excitation of a Rydberg atom
to a wave packet with large average principal
quantum number,
the ratio $l^\ast/n^\ast$ is small.
This means that the RSS we derive below
and the radial wave packets experimentally
constructed have corresponding classical
orbits that are highly elliptical,
with $e \simeq 1$.

\vglue 0.6cm
{\bf\noindent B. Oscillator Description of Radial Motion}
\vglue 0.4cm

We have found that
direct attempts to construct minimum-uncertainty
wave functions for the SQDT meet intractable
difficulties.
In fact,
the same is true of the pure hydrogenic case,
as has been known since the time of Schr\"odinger
\cite{schr}.
We sidestep the issue by
replacing $r$ and $p_r$
with a new set of classical variables,
$R$ and $P$,
in terms of which the
description of the motion is close to that
of a harmonic oscillator.
The new radial variable $R$ is chosen
to have a simple sinusoidal dependence
on the angle $\th$.
The resulting equations are relatively easy
to handle at the quantum level,
so that minimum-uncertainty packets can
be found analytically.
This elegant technique was originally introduced
in Ref.\ \cite{ni}
to construct `minimum-uncertainty coherent states'
for the Coulomb problem,
which correspond in the present context
to states with high angular momentum
(and small eccentricity).

For simplicity,
we pick $\th_0$ such that $f \th_0 = \frac {\pi} 2$.
The new radial variable is then given as
\beq
R \equiv \fr 1 r - \fr 1 {l^{\ast 2}} =
\fr e {l^{\ast 2}} \sin f \th
\quad .
\label{bigr}
\eeq
Its time derivative is
\beq
{\dot R} = \fr e {l^\ast r^2} \cos f \th
\quad ,
\label{bigrdot}
\eeq
and the corresponding conjugate momentum $P$ is
\beq
P \equiv - \fr {r^2} f {\dot R}
= - \fr {e l} {l^{\ast 2}} \cos f \th
\quad .
\label{bigp}
\eeq
The classical equations of motion can be written
\beq
{\dot R} = - \fr f {r^2} P
\quad ,
\label{cleq1}
\eeq
\beq
{\dot P} = \fr {l^2 f} {r^2} R
\quad .
\label{cleq2}
\eeq

The new classical variables satisfy the relations
\beq
\partial_r R =
\fr {\dot R} {\dot r} =
\fr 1 {l^\ast r} \sqrt { \fr
{ e^2 - l^{\ast 4} R^2}
{2 E r^2 + 2 r - l^{\ast 2} } }
\quad ,
\label{rel1}
\eeq
\beq
P = \fr 1 f {\dot r} = \fr {p_r} f
\quad .
\label{rel2}
\eeq
These can be used to reexpress
Eq.\ \rf{energy} as
\beq
\half P^2 + \half l^2 R^2 = \fr {e^2} {2 f^2}
\quad .
\label{newe}
\eeq
This equation has the form of
the energy equation for a simple harmonic oscillator
of frequency $l$ and energy $e^2/2f^2$.

\vglue 0.6cm
{\bf\noindent C. Derivation of Radial Squeezed States}
\vglue 0.4cm

We next pass to the quantum level.
The quantum radial hamiltonian for SQDT is
\beq
H = - \half \Bigl( \fr {d^2} {d r^2} + \fr 2 r \fr d {dr} \Bigr)
- \fr 1 {r} + \fr {{l^\ast} ( {l^\ast} + 1)} {2 r^2}
\quad .
\label{hqm}
\eeq
The canonical radial coordinate $r$ and its conjugate momentum
$p_r = - i ( \partial_r + \fr 1 r )$
obey the commutation relation $[r,p_r] = i$.
The eigensolutions for this hamiltonian are given in
Eq.\ \rf{rdef}.

The quantum operators corresponding to the new classical
variables $R$, $P$ are
\beq
R = \fr 1 r - \fr 1 {{l^\ast} ( {l^\ast} + 1)}
\quad ,
\label{bigrqm}
\eeq
\beq
P = \fr {p_r} f = - \fr i f ( \partial_r + \fr 1 r )
\quad .
\label{bigpqm}
\eeq
They obey the commutation relation
\beq
[R,P] =
- \fr i f \fr 1 {r^2}
\quad .
\label{rpcom}
\eeq
The presence of the factor $1/r^2$ in this equation
reflects the unconventional choice of coordinate $R$
and shows that,
despite the similarities described above,
the system is not a true quantum-mechanical
simple harmonic oscillator.

The uncertainty product $\De R \De P$
follows from Eq.\ \rf{rpcom} and is given by
\beq
\De R \De P \ge \fr 1  {2 f} \vev{\fr 1 {r^2}}
\quad .
\label{uncrp}
\eeq
At a fixed time,
the minimum-uncertainty wave functions
therefore satisfy
\beq
\bigl( R - \vev{R} \bigr) \psi =
i A \bigl( P - \vev{P} \bigr) \psi
\quad ,
\label{minrel1}
\eeq
where
\beq
A = \fr {2 f (\De R )^2}{\vev{\fr 1 {r^2}}}
= \fr {\De R} {\De P}
\quad
\label{bdef}
\eeq
is a real constant.
Defining the parameters
\beq
\al = \fr f A - 1 ~~,~~~~
\ga_0 = \fr f A \vev{\fr 1 r}~~,~~~~
\ga_1 =  - \vev{p_r}
\quad ,
\label{igdef}
\eeq
equation \rf{minrel1} becomes
\beq
\left( \partial_r - \fr {\al} r
+ ( \ga_0 + i \ga_1 ) \right) \psi (r) = 0
\quad .
\label{minrel3}
\eeq

The solution to Eq.\ \rf{minrel3} is
\beq
\psi (r) = N r^{\al} e^{-\ga_0 r} e^{-i \ga_1 r}
\quad ,
\label{ras}
\eeq
where $N$ is a normalization constant.
These states form a three-parameter family of
squeezed states
minimizing the uncertainty relation \rf{uncrp}
at a fixed time.
These are our radial squeezed states (RSS).

Our derivation has demonstrated that,
in any atom for which SQDT provides a good description,
the RSS are candidates for Rydberg wave packets.
Moreover,
they have a relatively simple analytical form.
In particular,
these results apply
to the alkali-metal atoms rubidium and potassium,
for which the experiments on radial wave packets
have been done to date.

For the special choice $\al = l$,
corresponding to RSS with uncertainty ratio
$A$ equal to that of the ground state,
the RSS specialize from the three-parameter family
of squeezed states to a two-parameter family of
coherent states
\cite{nissfoot}.
They then provide SQDT versions of the
`minimum-uncertainty coherent states'
for three-dimensional systems that were
introduced in
Ref.\ \cite{ni}.
However, they correspond to states of Rydberg atoms
with high angular momentum.
This means they are unsatisfactory as descriptions of
Rydberg atoms prepared by excitation with
a single short laser pulse
with no external field present,
since these have large $n$ and $l=1$.
We therefore do not discuss these coherent states further
in the present work.

\vglue 0.6cm
{\bf\noindent D. Properties of Radial Squeezed States}
\vglue 0.4cm

In this section,
we present some time-independent
properties of the RSS.
The time-dependent behavior is discussed in
Sec.\ V.

First,
we normalize $\ps (r)$.
Imposing
\beq
\int_0^\infty r^2 |\psi (r)|^2 dr = 1
\quad
\label{norm}
\eeq
gives
\beq
N = \fr {(2 \ga_0 )^{2 \al + 3}} {\Ga (2 \al + 3)}
\quad .
\label{bign}
\eeq
This normalization requires $\al \ge -1$
and $\ga_0 > 0$.

Certain expectation values are of particular use
in the description of the RSS and the comparison
with other results.
Here are a few key expectations,
all calculated with normalized RSS.
We find
\beq
\vev{r} = \fr {2 \al + 3} {2 \ga_0} ~~,~~~~
\vev{\fr 1 r} = \fr {\ga_0} {\al + 1}
\quad ,
\label{exp2}
\eeq
\beq
\vev{r^2} = \fr {(\al + 2)(2 \al + 3)} {2 {\ga_0}^2} ~~,~~~~
\vev{\fr 1 {r^2}} = \fr {2 {\ga_0}^2} {(\al + 1)(2 \al + 1)}
\quad ,
\label{exp4}
\eeq
\beq
\vev{p_r} = - \ga_1~~,~~~~
\vev{{p_r}^2} = \fr {{\ga_0}^2} {2 \al + 1} + {\ga_1}^2
\quad .
\label{exp6}
\eeq
It follows that the expectation value of the hamiltonian,
which is the energy expectation value of the RSS,
is given by
\beq
\vev{H} =
\fr {\ga_0^2} {2 (\al + 1)(2 \al + 1)}
\left( 2 \ga_0^2 l^\ast (l^\ast + 1) + \al + 1 \right)
- \fr {\ga_0} {\al + 1}
+ \fr {{\ga_1}^2} 2
\quad .
\label{exp7}
\eeq
{}From these expressions,
it is apparent that
the parameter $\al$ is
further constrained by the requirement
that the kinetic and potential energies of the
RSS be separately normalizable,
so that $\al > -\half$.

It is also of interest to determine the uncertainties
in the conventional coordinates $r$ and $p_r$.
We find
\beq
\De r = \fr {\sqrt{2 \al + 3}} {2 \ga_0}~~,~~~~
\De p_r = \fr {\ga_0} {\sqrt{2 \al + 1}}
\quad .
\label{uncpr}
\eeq
This implies that the RSS obey the uncertainty relation
\beq
\De r \De p_r = \half \sqrt{\fr {2 \al + 3} { 2 \al + 1}}
\quad .
\label{uncal}
\eeq
As might be expected,
the RSS are \it not \rm minimum-uncertainty states
in the usual variables $r$ and $p_r$,
although by construction they satisfy the
equality in the minimum-uncertainty relation
\rf{uncrp}.

The possible value of the uncertainty product
ranges from an arbitrarily large value
when $\al$ is near its minimum of $-\half$,
through  $\De r \De p_r = \sqrt 3/2\simeq 0.866$
when $\al = 0$,
to values falling as $\De r \De p_r \approx \half (1+ 1/\al)$
for large $\al$.
As is discussed in Sec.\ V,
large values of $\al$ are required
to match key features of the RSS with the
experimentally produced Rydberg wave packets
at the first pass through the
outer apsidal point of the orbit.
We therefore expect the appropriate RSS to have
uncertainty product close to $\half$ at that time.
This feature agrees
with the behavior of Rydberg atoms excited by a short laser pulse.
Numerical calculations performed for hydrogen
show that $\De r \De p_r$ is close to its minimum
when the electron is near the outer turning point
\cite{pasa,pasb}.

The radial probability distribution,
\beq
f(r) \equiv r^2 |\psi (r)|^2
= N^2 r^{2 \al + 2} e^{-2 \ga_o r}
\quad ,
\label{dist}
\eeq
determines the shape of the RSS.
The function $f(r)$
is asymmetrical in $r$.
Its central maximum is located at $r=r_0$,
where
\beq
r_0 = \fr {\al + 1} {\ga_0} = \fr 1 {\vev{\fr 1 r}}
\quad .
\label{cent}
\eeq
The curvature of the envelope at $r_0$
can conveniently be measured by the ratio
\beq
C_0 = \fr {f^{\prime\prime}(r_0)}{f(r_0)}
= - \fr {2 \ga_0^2}{\al +1}
\quad .
\label{curv}
\eeq
The specification of the RSS distribution is complete
if $r_0$ and $C_0$ are known.

The RSS envelope $f(r)$
is a gamma distribution.
The mean is just $\vev r$,
while the variance is $(\De r)^2$.
These are given in Eqs.\ \rf{exp2}
and \rf{uncpr}.
Other standard measures of the asymmetry of the
function $f(r)$,
such as the momental skewness or the kurtosis,
can be obtained as simple expressions
in terms of $\al$ and $\ga_0$.
The detailed shape of the distribution
is determined by its moments $m_k$ about the mean,
defined as
\beq
m_k = \int_0^{\infty}~(r-\vev r )^k~f(r)~r^2~dr
\quad .
\label{moments}
\eeq
These can be found by direct integration
or as the coefficients of the power-series
expansion of a relatively
simple moment-generating function.
In principle,
the moments $m_k$ provide a basis for a detailed
comparison of the RSS with wave packets
obtained from other theory or experiment.
We do not further pursue this line of inquiry here.

\vglue 0.6cm
{\bf\noindent V. TIME EVOLUTION OF RADIAL SQUEEZED STATES}
\vglue 0.4cm

At this stage,
we have identified the RSS as candidate
Rydberg wave packets
and have discussed several time-independent features.
In this section,
we discuss some time-dependent properties,
in particular the RSS time evolution and autocorrelation
functions.
We also compare our results for the RSS with
results presented elsewhere in the literature.

\vglue 0.6cm
{\bf\noindent A. Time Evolution: Theory}
\vglue 0.4cm

Since the RSS have a relatively simple form,
their behavior as a function of time
can be studied both numerically and analytically.
We have done both.
Our numerical integrations are performed
using the Crank-Nicholson method.
Our analytical studies apply to the time
evolution of RSS and
to their autocorrelation functions,
including issues such as
characterizing full and fractional revivals
as a function of the
supersymmetry integer $I(l)$
and the quantum defect $\de (l)$.
This subsection presents
some analysis for time-dependent properties.
In what follows,
we define the origin $t=0$ to be the time
that the packet first reaches
the outer apsidal point.

The full three-dimensional
time-evolved wave packet can be written as a sum over
the complete set of SQDT states:
\beq
\Psi ({\vec r},t) = Y_{1 0} (\th,\ph) \sum_n c_n
R_{ {n^\ast} {l^\ast} } (r)
e^{-i E_{n^\ast} t}
\quad ,
\label{timeexps}
\eeq
where $l^\ast = 1 - \de (1) + I(1)$ is specified for p states.
In principle,
the expansion includes the continuum states.
However,
for cases of interest the contributions
from continuum states are negligible.
The coefficients $c_n$ are determined by
requiring that the initial wave function
$\Ps (\vec r, 0)$
has radial piece given by the RSS $\psi (r)$ of
Eq.\ \rf{ras},
i.e.,
$\Ps (\vec r, 0) = Y_{10}(\th,\ph) \ps (r)$.
This implies
\beq
c_n = \bigl< R_{ {n^\ast} {l^\ast} } (r) \bigm| \psi (r) \bigr>
\quad .
\label{cn}
\eeq
Using the expression for $R_{ {n^\ast} {l^\ast} } (r)$ in
Eq.\ \rf{rdef}
and performing the integration,
we get after some calculation
the result
$$c_n =
N \fr {\Ga (\al + l^\ast + 3)} {\Ga (l^\ast + 1)}
\left( \fr {\Ga (n^\ast + l^\ast + 1)}
{\Ga (n^\ast - l^\ast)}\right)^\half
\fr {n^{\ast (\al + 1)}}
{\left[ n^\ast (\ga_0 + i \ga_1) + 1 \right]^{\al + l^\ast + 1}}
$$
\beq
\times {_2}F{_1} \left( {l^\ast} + 1 - {n^\ast}, \al + {l^\ast} + 3;
2 {l^\ast} + 2; \fr 2 {n^\ast (\ga_0 + i \ga_1) + 1} \right)
\quad .
\label{mess2}
\eeq
Here,
${_2}F{_1}$ is a hypergeometric function,
and $N$ is the normalization constant given in
Eq.\ \rf{bign}.

The coefficients in Eq.\ \rf{mess2}
can also be found via a suitable integration over
the analytical propagator for the Coulomb problem
obtained recently in Ref.\ \cite{blin}.
For the examples of interest,
we find that the coefficients in the sum
over states in Eq.\ \rf{timeexps} are strongly
peaked around a central value ${\bar n}$ of $n$.
Accurate numerical approximations can be
obtained by truncating the sum to a subset of
states centered on ${\bar n}$.

Another quantity of interest for the characterization
of the time evolution is
the absolute square of the autocorrelation function,
$A(t) = {\bigm| \bigl< \Psi (t) \bigm| \Psi (0) \bigr> \bigm|}^2$.
In terms of the coefficients in Eq.\ \rf{mess2},
it is given as
\beq
A(t) = {\Bigm| \sum_n \bigm| c_n
\bigm|^2 e^{-i E_{n^\ast} t} \Bigm|}^2
\quad .
\label{biga}
\eeq

Studies of the behavior of a hydrogenic radial packet
as a function of time
show that wavefunction dispersion is followed by
the appearance of revivals
\cite{pasa,avp,naa}.
We have repeated these analyses
for the RSS within the context of the SQDT.
As the RSS evolves in time
from its initial configuration at $t=0$,
it disperses.
However,
it is a trapped object,
being bound to the atomic core.
The dispersion therefore eventually leads
to a situation in which what were initially
the front and back parts of the packet interfere.
This provides one signal for the eventual decoherence
of the packet.
For an RSS with eigenstate decomposition
peaked about a principal quantum number $\bar n^\ast$
and dominantly contained within a range $\de n^\ast$,
we find that in the SQDT that the
interference time
$t_{\rm int}^\ast$ at which the packet effectively collapses
is given approximately by the expression
\beq
t_{\rm int}^\ast
= \frac 13 \fr {\bar n^\ast}{\de n^\ast} T^\ast_{\rm cl}
\quad ,
\label{tint}
\eeq
where $T_{\rm cl}^\ast$ is given by Eq.\ \rf{tclqd2}.

At a time substantially later than $t_{\rm int}$,
the RSS reforms approximately into its original shape.
For the SQDT,
we find that this behavior occurs
near a revival time $t_{\rm rev}^\ast$ given by
\beq
t_{\rm rev}^\ast = \frac 13 {\bar n^\ast} T^\ast_{\rm cl}
\quad .
\label{trev}
\eeq
The new wave packet oscillates with periodicity
equal to the classical orbital period $T_{\rm cl}^\ast$.
We refer to this packet as a full revival
\cite{footrev}.

At certain times between
$t_{\rm int}^\ast$ and
$t_{\rm rev}^\ast$,
the RSS gathers into $r$ spatially separated packets
called fractional revivals.
A full analysis in the SQDT,
along the lines of that presented in
Ref.\ \cite{avp},
suggests that these times are given by
rational fractions $2p/q$
of the revival time $t_{\rm rev}^\ast$,
where $p$ and $q$ are relatively prime
and where the number of packets is $r=q$ if $q$ is odd
or $r= q/2$ if $q$ is even.
For the purposes of the discussions in the subsections below,
we single out
the particular fractional-revival times
\beq
t_{r}^\ast = \frac 1r t_{\rm rev}^\ast
\quad .
\label{tfrac}
\eeq
The wave-function period of these fractional revivals
in the SQDT is given by
\beq
T_r^\ast = \frac 1r T^\ast_{\rm cl}
\quad .
\label{perfrac}
\eeq
With our conventions,
the full revival corresponds to the case $r=1$.

\vglue 0.6cm
{\bf\noindent B. Initialization}
\vglue 0.4cm

To compare the properties of the RSS
for different alkali-metal atoms
with other theory and experiment,
we need a procedure for determining the values
of the parameters $\al$, $\ga_0$, and $\ga_1$
in a given situation.
In a Rydberg atom that has been excited by a short
laser pulse,
the uncertainty product
$\De r \De p_r$
is expected to be lowest at the first pass through
the outer turning point
\cite{pasa,pasb}.
We therefore take this as our initialization point,
and, as above,
define it as the time origin $t=0$.

Fixing the initial form of the RSS
requires the specification of three quantities.
We choose these as the physical quantities
$\vev{p_r}$,
$\vev{r}$,
and $\vev{H}$.
The natural choice for the expectation value
of the radial momentum at the apsidal point
is zero.
Similarly,
a natural choice for the expectation value of
$r$ at the initialization point is
the outer apsidal point of the orbit,
\beq
r_{\rm out}^\ast = n^{\ast 2} \left[ 1 + \sqrt{1 -
\fr {{l^\ast}({l^\ast} + 1)} {n^{\ast 2}}} ~\right]
\quad .
\label{rclqm}
\eeq
Note that this differs from the outer apsidal
point of the classical orbit,
given in Eq.\ \rf{rcl},
by the use of the quantum-mechanical eigenvalue
for the angular-momentum operator.
Finally,
one natural choice for the energy expectation value
is the energy
$E_{\bar n^\ast} = - 1/2 \bar n^{\ast 2}$
of the central state $\bar n$
in the range of states
excited by the short laser pulse.

Thus,
we initialize the RSS
by choosing the parameters $\al$, $\ga_0$, $\ga_1$
to satisfy the following equations:
\beq
\vev{p_r} = 0
\quad ,
\label{cond1}
\eeq
\beq
\vev{r} = r_{\rm out}^\ast
\quad ,
\label{cond2}
\eeq
\beq
\vev{H} = E_{\bar n^\ast}
\quad .
\label{cond3}
\eeq
The expectation values appearing in these equations
are given in Eqs.\ \rf{exp2}, \rf{exp6}, and \rf{exp7}.
The first condition,
Eq.\ \rf{cond1},
can be inverted immediately to give $\ga_1 =0$.
This leaves two conditions to fix the two remaining
parameters $\al$ and $\ga_0$ in terms of the two quantities
$n^\ast$ and $l^\ast$.
These in turn depend on the two quantities
$I(l)$ and $\de (l)$.

For the Rydberg wave packets obtained in experiment
the value of $l^\ast$ is much smaller than the value
of $n^\ast$.
This means that the value of $\al$ is much greater
than that of $\ga_0$.
In the cases we consider below,
$\al$ is large and exceeds $\ga_0^2$
by more than six orders of magnitude.
{}From Eqs.\ \rf{uncpr}
it follows that the initial squeezing in $p_r$
exceeds that in $r$ by about six orders of magnitude:
$\De r /\De p_r \sim \al/\ga_0^2 \sim 10^6$.
For more details about the uncertainty properties of the RSS,
including a discussion of the time-dependence of the
uncertainty product and ratio,
see Ref.\ \cite{bkrss}.

\vglue 0.6cm
{\bf\noindent C. Examples: Time Evolution}
\vglue 0.4cm

In this subsection,
we discuss two examples of the time evolution
of RSS.
The first is an RSS for hydrogen,
with dominant energy component corresponding
to principal quantum number
$\bar n=85$.
This example serves as a basis
for comparison with other theoretical results.
The second example is for rubidium.
To determine the effects of the quantum defect on
the time evolution,
we choose for this case the same value of $\bar n$,
which corresponds to $\bar n^\ast \simeq 82.35$.

For our first example,
the SQDT reduces to the usual Coulomb case for p states:
$I(1) = \de (1) = 0$,
$l^\ast = l = 1$,
and $\bar n^\ast = \bar n = 85$.
{}From the conditions
\rf{cond2}
and \rf{cond3},
we determine that the RSS parameters are
$\al \simeq 168.225$
and $\ga_0 \simeq 0.0117465$.
This gives an initial uncertainty product of
$\De r \De p_r \simeq 0.50148$.
We find that initially the wave
packet moves toward the inner turning point.
The packet decreases in amplitude,
broadens,
and starts to oscillate as it approaches
the origin.
After the bounce,
it regathers into a coherent packet as it
approaches the outer turning point again.
We find that the periodic motion of the RSS is
consistent with the classical orbital period
$T_{\rm cl}^\ast =T_{\rm cl}
= 2 \pi \bar n^3 \simeq 93.3$ psec.

Figures 3 and 4 show the envelope of the RSS
at different times during the first orbit or so.
In Fig.\ 3,
the envelope of the RSS at $t=0$ is centered about the
outer turning point near
$r_{\rm out} \approx 2 \bar n^2 \simeq 14450$ a.u.
The direction of propagation is towards the origin.
The envelopes at times $t\simeq 19.4$ and $38.7$ psec
are also shown.
Fig.\ 4
shows the RSS envelopes after reflection off the core,
while approaching and passing through the outer
turning point again.
{}From left to right,
the graphs are at times $t\simeq 58.1$, $77.4$ and $96.8$ psec.
The latter plot displays a decreased amplitude
because the packet is a few picoseconds
into its second orbit and is therefore moving towards the origin
again.

The generic behavior we find is consistent with the results
of our perturbative calculation shown in
Fig.\ 1.
However,
the perturbative calculation starts to break down after
about one orbit,
and comparisons cannot be made for long times.
Our results compare well with
Fig.\ 1(b)
of
Ref.\ \cite{pasa},
which shows the Rydberg wave packet during the second
part of its orbit,
obtained by numerically
integrating the time-dependent Schr\"odinger equation
for an electron in hydrogen subject to a short laser pulse.
This comparison
suggests that the RSS
are good models for radial wave packets in hydrogen.

Our second example involves rubidium p states
with $\bar n = 85$,
which have $\de (1) \simeq 2.65$,
$I(1) = 3$,
$l^\ast \simeq 1.35$,
and $n^\ast \simeq 82.35$.
These values determine the RSS parameters as
$\al \simeq 162.91$
and $\ga_0 \simeq 0.0121233$,
with an initial uncertainty product
$\De r \De p_r \simeq 0.50153$.

Figures 5 and 6
show the time evolution of the envelope
of this RSS for rubidium.
The time intervals in
Figs. 5 and 6
are the same as those used for hydrogen in
Figs. 3 and 4.
Some differences between the rubidium and hydrogen
examples are apparent.
These may largely be attributed
to the quantum defect.
For example,
the outer apsidal point for rubidium is
$r_{\rm out} \approx 2 \bar n^{\ast 2} \simeq 13560$ a.u.
and is closer to the core
than the corresponding point for hydrogen.
Also,
comparison of Figs. 3 and 5
shows that at $t \simeq 38.7$ psec
the rubidium RSS is oscillating significantly more than
the hydrogen one.
This is a consequence of the reduction of the
classical orbital period of rubidium,
$T_{\rm cl}^\ast = 2 \pi \bar n^{\ast 3} \simeq 84.9$ psec,
relative to that of hydrogen.
At $38.7$ psec,
the rubidium RSS is closer to halfway through
its orbit and hence exhibits more quantum-like behavior.

\vglue 0.6cm
{\bf\noindent D. Examples: Autocorrelation Functions}
\vglue 0.4cm

In this subsection,
we provide some examples of autocorrelation functions
for RSS.
Three cases are considered.
The first two use the same RSS for hydrogen and rubidium
considered in the previous subsection.
The third example we use is an RSS for potassium
with $\bar n = 67$,
which is convenient for comparison with experiment.

Our first example is the RSS for hydrogen with $\bar n = 85$.
The associated parameters $\al$ and $\ga_0$
are given in the previous subsection.
The discussion in Sec.\ V A suggests that
the qualitative features of the motion are governed
by the classical orbital period
$T_{\rm cl}\simeq 93.3$ psec,
by the interference time
$t_{\rm int} \simeq 4 T_{\rm cl}$
obtained with a spread $\de n \sim 7$,
and by the revival time
$t_{\rm rev} \simeq 2.6$ nsec.

Figure 7
shows the absolute square of the autocorrelation function
for hydrogen with $\bar n=85$.
Near $t=0$ the autocorrelation
exhibits peaks spaced by the classical orbital period,
reflecting the oscillatory motion between the outer
and inner apsidal points expected at early times.
About four orbits can be distinguished before the
packet collapses.
At times near the predicted value of $t_{\rm rev}$,
peaks spaced by the classical orbital period reappear.
The figure also reveals the appearance of
fractional revivals corresponding to the values
$r = 4$, 3, and 2
at times
$t_4 \simeq 0.65$ nsec,
$t_3 \simeq 0.88$ nsec,
and $t_2 \simeq 1.3$ nsec,
in agreement with the theory.
As expected,
these fractional revivals exhibit wave-function periodicities
of
$T_4 \simeq 23.3$ psec,
$T_3 \simeq 31.1$ psec,
and $T_2 \simeq 46.7$ psec,
respectively.
Note that the maximum values of $A$ near the times $t=t_r$
decrease with increasing $r$,
as might be expected from the spatial separations and numbers
of the fractional revivals.

Our next example is the RSS for rubidium
with $\bar n =85$.
As described in the previous subsection,
this corresponds to $\bar n^\ast \simeq 82.35$.
The SQDT predicts a modified behavior of this RSS,
with key features determined
by the classical orbital period
$T_{\rm cl}^\ast\simeq 84.9$ psec,
by the interference time
$t_{\rm int}^\ast \simeq 4 T_{\rm cl}^\ast$,
and by the revival time
$t_{\rm rev}^\ast \simeq 2.3$ nsec.
Note that although the interference time
is different from that in hydrogen,
it remains given by about four classical orbital periods
because the values of $\bar n$ and $\de n$
do not change significantly; cf.\ Eq.\ \rf{tint}.

Figure 8
shows the absolute square of the autocorrelation function
for this example.
The overall behavior is similar to that of hydrogen.
In agreement with the predicted interference time,
about four orbits occur before wave-packet collapse.
Larger correlations separated
by the classical orbital period reappear near the
theoretical value of $t_{\rm rev}^\ast$.
Fractional revivals with
$r = 4$, 3, and 2
can be seen near
$t_4^\ast \simeq 0.58$ nsec,
$t_3^\ast \simeq 0.78$ nsec,
and $t_2^\ast \simeq 1.2$ nsec,
as expected.
The associated periodicities
are
$T_4^\ast \simeq 21.2$ psec,
$T_3^\ast \simeq 28.3$ psec,
and $T_2^\ast \simeq 42.5$ psec.

Our last example is for the RSS in potassium
with $\bar n = 67$.
The associated p states have $\de (1) \simeq 1.71$,
$I(1) = 2$,
$l^\ast \simeq 1.29$,
and $n^\ast \simeq 65.29$.
The corresponding RSS parameters are
$\al \simeq 120.306$
and $\ga_0 \simeq 0.0142822$.
The initial uncertainty product is
$\De r \De p_r \simeq 0.50207$.
For this case,
the classical orbital period is
$T_{\rm cl}^\ast\simeq 42.3$ psec.
With $\de n^\ast \sim 7$ still,
the interference time is reduced to
$t_{\rm int}^\ast \simeq 3 T_{\rm cl}^\ast$,
i.e., about three orbits are expected
to occur before the packet collapses.
The revival time
is now significantly smaller,
$t_{\rm rev}^\ast \simeq 0.92$ nsec.

The absolute square of the autocorrelation function for
this potassium RSS is shown in Fig.\ 9.
Once again,
the features agree with the theory.
Only the fractional revivals with
$r = 3$ and 2 can be clearly seen, however.
They occur near
$t_3^\ast \simeq 0.31$ nsec
and $t_2^\ast \simeq 0.46$ nsec,
as predicted.
The wave-function periodicities are
$T_3^\ast \simeq 14.1$ psec
and $T_2^\ast \simeq 21.2$ psec,
respectively.

This example can be compared to the experiment presented
in Ref.\ \cite{yms},
in which potassium radial wave packets
were excited in a pump-probe experiment.
Oscillations in the photoion signal were observed that
are consistent with $T_{\rm cl}^\ast$ and $t_{\rm rev}^\ast$
calculated from SQDT for a potassium RSS
with $\bar n^\ast \simeq 65$.
The revival peaks observed would correspond
to the autocorrelation peaks exhibited in
Fig.\ 9 starting at about $0.8$ nsec.
Similarly,
examination of the experimental results presented in
Refs.\ \cite{yes,mmhe} concerning the observation
of fractional revivals with $r=2$
reveals behavior consistent with the analysis above.

\vglue 0.6cm
{\bf\noindent VI. SUMMARY AND CONCLUSIONS}
\vglue 0.4cm

In this paper,
we have investigated two possibilites for
describing Rydberg wave packets obtained by
atomic excitation with short laser pulses.
The approaches are somewhat complementary
and both are developed in the context of SQDT,
which can be viewed as an analytical model for
single-body effective wave functions of Rydberg atoms.
As such, the results we obtain
are relevant both for hydrogen
and for non-hydrogenic systems,
including in particular the alkali-metal atoms
used in experiments.

The first approach is a direct extension of
time-dependent perturbation theory for the
generation and evolution of the wave packets.
The success of this approach suggests that
the SQDT may be of use in calculations of this type
carried out for alkali-metal atoms.

We have concentrated primarily on the second approach,
which is an attempt to describe Rydberg wave packets
in terms of a class of analytical squeezed states,
the RSS.
Using the SQDT,
we succeeded in constructing the RSS
as a three-parameter family of analytical states
minimizing an uncertainty relation
in nonstandard coordinates.
The analytical form
of the RSS,
which are gamma distributions,
means that relatively simple and/or
analytical expressions can be obtained for
various properties.
The use of SQDT ensures that the RSS
provide candidates for the analytical description
of Rydberg wave packets in atoms other than hydrogen.

In terms of the radial coordinates $r$ and $p_r$,
the RSS have close to minimum uncertainty
for the cases we consider.
To match with other theory and experiment
in specific cases,
we initialize the RSS
at the outer apsidal point of the orbit.
The three parameters are fixed by matching the
expectation values of the position, momentum, and
energy at that point.
The ensuing behavior is completely determined.
We derive relatively simple expressions for the time-evolved
RSS and for their autocorrelation functions.
We also predict the dependence of
characteristic features of the behavior
on the quantum defects.
These include
the initial orbital period,
the time of packet collapse,
and the full and fractional revival times
and their associated orbital periodicities.

We have presented examples of time evolutions
and autocorrelation functions both for hydrogen
and for alkali-metal atoms.
The behavior agrees with our theoretical expectations.
The properties and behavior of the RSS are
also in agreement with results on
radial probability distributions,
time evolution,
and autocorrelation functions
obtained by other authors for Rydberg wave packets.
It appears that the RSS provide an analytical tool
for describing and distinguishing
experimental and theoretical features of
Rydberg-atom radial wave packets
produced by short-pulsed laser fields.

\newpage

{\bf\noindent ACKNOWLEDGMENTS}
\vglue 0.4cm

We enjoyed conversations with Charlie Conover
and Duncan Tate.
R.B. thanks Colby College for a Science Division Grant.
Part of this work was performed while V.A.K. was
visiting the Aspen Center for Physics.

\vglue 0.6cm
{\bf\noindent REFERENCES}
\vglue 0.4cm

\vfill\eject

\baselineskip=16pt
{\bf\noindent Figure Captions}
\vglue 0.4cm

\begin{description}

\item[{\rm Fig.\ 1: }]
Hydrogen radial wave packet with $\bar n=85$.
The unnormalized radial probability distribution
$f(r)$ is plotted as
a function of radial distance $r$ in a.u.
{}From left to right,
the graphs are shown for the times
$t= \frac 1 9 T_{\rm cl}$, $\frac 2 9 T_{\rm cl}$,
$\frac 3 9 T_{\rm cl}$, and
$\frac 4 9 T_{\rm cl}$, where $T_{\rm cl} = 93.3$ psec.

\item[{\rm Fig.\ 2: }]
Rubidium radial wave packet with $\bar n=85$.
The unnormalized radial probability distribution
$f(r)$ is plotted as
a function of radial distance $r$ in a.u.
{}From left to right,
the graphs are shown at the same times as in Fig.\ 1:
$t=10.4$, $20.7$, $31.1$, and $41.5$ psec.

\item[{\rm Fig.\ 3: }]
Hydrogen RSS with $\bar n=85$:
first part of the orbital motion.
The unnormalized radial probability distribution
$f(r)$ is plotted as
a function of radial distance $r$ in a.u.
{}From right to left,
the graphs are shown
at the times $t=0$, $19.4$, and $38.7$ psec.

\item[{\rm Fig.\ 4: }]
Hydrogen RSS with $\bar n=85$:
second part of the orbital motion.
The unnormalized radial probability distribution
$f(r)$ is plotted as
a function of radial distance $r$ in a.u.
{}From left to right,
the graphs are shown
at the times $t=58.1$, $77.4$, and $96.8$ psec.

\item[{\rm Fig.\ 5: }]
Rubidium RSS with $\bar n=85$:
first part of the orbital motion.
The unnormalized radial probability distribution
$f(r)$ is plotted as
a function of radial distance $r$ in a.u.
{}From right to left,
the graphs are shown
at the same times as in Fig.\ 3.

\item[{\rm Fig.\ 6: }]
Rubidium RSS with $\bar n=85$:
second part of the orbital motion.
The unnormalized radial probability distribution
$f(r)$ is plotted as
a function of radial distance $r$ in a.u.
{}From left to right,
the graphs are shown
at the same times as in Fig.\ 4.

\item[{\rm Fig.\ 7: }]
Absolute square of the
autocorrelation function for hydrogen RSS with $\bar n = 85$
as a function of time in nanoseconds.

\item[{\rm Fig.\ 8: }]
Absolute square of the
autocorrelation function for rubidium RSS with $\bar n = 85$
as a function of time in nanoseconds.

\item[{\rm Fig.\ 9: }]
Absolute square of the
autocorrelation function for potassium RSS with $\bar n = 67$
as a function of time in nanoseconds.

\end{description}


\begin{thebibliography}{xx}
\bibitem{nissfoot}
We remark that similar
squeezed states for the pure Coulomb problem
were obtained as part of the method of derivation
of the `minimum-uncertainty coherent states'
in Ref.\ \cite{ni}.
Although not discussed as such in those papers,
the role of these objects
as squeezed states was subsequently
pointed out in Ref.\ \cite{niss}.
\bibitem{footrev}
The reader is warned that for the hydrogenic case
some difference
in nomenclature exists in the literature,
with this new packet
often called a `one-half revival.'
The full revival time then may be taken as
the quantity equivalent to $2t_{\rm rev}^\ast$,
although evidently with zero supersymmetry integer $I(l)$
and zero quantum defect $\de (l)$.
\end{thebibliography}
\end{document}